# Experimental evidence on the dissipationless transport of chiral edge state of the high-field Chern insulator in MnBi$_2$Te$_4$ nanodevices

Zhe Ying[1,a], Shuai Zhang[1,a,*], Bo Chen[1,a], Bin Jia[1], Fucong Fei[1,*], Minhao Zhang[1], Haijun Zhang[1], Xuefeng Wang[2], and Fengqi Song[1,*]

[1]National Laboratory of Solid State Microstructures, Collaborative Innovation Center of Advanced Microstructures, and School of Physics, Nanjing University, Nanjing 210093, China

[2]National Laboratory of Solid State Microstructures, Collaborative Innovation Center of Advanced Microstructures, and School of Electronic Science and Engineering, Nanjing University, Nanjing 210093, China

[a] These authors contributed equally.
[*]Corresponding authors. Email: S.Z.(szhang@nju.edu.cn), F.F.(feifucong@nju.edu.cn), and F.S.(songfengqi@nju.edu.cn)

**Abstract**

We demonstrate the dissipationless transport of the chiral edge state (CES) in the nanodevices of quantum anomalous Hall insulator candidate $MnBi_2Te_4$. The device presents a near-zero longitudinal resistance together with a quantized Hall plateau in excess of 0.97 $h/e^2$ over a range of temperatures from very low up to the Néel temperature of 22 K. Each of four-probe nonlocal measurements gives near-zero resistance and two-probe measurements exhibit a plateau of +1 $h/e^2$, while the results of three-probe nonlocal measurements depend on the magnetic field. This indicates non-dissipation as well as the chirality of the edge state. The CES shows three regimes of temperature dependence, i.e., well-preserved dissipationless transport below 6 K, variable range hopping while increasing the temperature and thermal activation at higher than 22 K. Even at the lowest temperature, a current of over 1.4 μA breaks the dissipationless transport. These form a complete set of evidences of the Chern insulator state in the $MnBi_2Te_4$ systems.

**Introduction**

The Hall resistance can be quantized without the formation of the Landau level (LL), which is the case of quantum anomalous Hall effect (QAHE) and also known as the Chern insulator state [1]. Driven by the coexistence of nontrivial topology of the electronic structure and ferromagnetism exchange, the long-desired QAHE was first observed in the magnetically doped topological insulator (TI) Cr-(Bi,Sb)$_2$Te$_3$ [2]. However, inhomogeneity due to the presence of magnetic dopants may lead to a rather small effective magnetic exchange gap, and thus the QAHE onset temperature in a magnetically doped TI is still well below 2 K, despite considerable efforts have been devoted to increasing this [3-5]. MnBi$_2$Te$_4$ was predicted and experimentally verified to be an intrinsic magnetic TI [6-15], where the magnetism originates from long-range ordered Mn atoms, rather than from stochastically distributed dopants in previous systems. Thus, it becomes a promising platform to allow QAHE and Chern insulator states at higher temperature, as previously observed in recent experiments in flakes of MnBi$_2$Te$_4$ [16-20]. Even the interesting axion insulator and high Chern number state can be simulated [17, 18]. Research enthusiasm is ignited afterwards on the family of MnBi$_2$Te$_4$.

Besides the quantized Hall plateaus, dissipationless chiral edge state (CES) should be demonstrated to form a complete set of evidences of Chern insulator states. This is also unique in the future device applications of a Chern insulator device. However, it has been rarely studied and still lack particularly in satisfactory devices with consistent temperature dependence between the CES and Hall plateau.

In this study, we demonstrate the dissipationless transport of the CES in the high-field Chern insulator state of the MnBi$_2$Te$_4$ device. The precise quantization of the Chern insulator is observed with a longitudinal resistance of ~ 0.0002 $h/e^2$ and a Hall plateau of ~ -1.005 $h/e^2$, where $h$ is the Planck constant and $e$ is the electron charge. The well-defined quantization can survive up to 22 K, with the plateau being in excess of 0.97 $h/e^2$, which allows systematic study of the dissipationless transport of the CES. By considering both the local and nonlocal transport, we studied the temperature-dependent scaling behaviors. The transition from well-preserved dissipationless state to variable range hopping and thermal activation is found with increasing temperature. We also investigate the current-induced breakdown of the dissipationless CES.

**Results**

**Evidences of the high-magnetic-field Chern insulator state.**

The MnBi$_2$Te$_4$ crystal was grown using the flux method [13]. After mechanical exfoliation and microfabrication, a MnBi$_2$Te$_4$ field-effect-transistor was obtained [21]. A hexagonal boron nitride (h-BN) flake was then transferred to the top of the MnBi$_2$Te$_4$ to protect the sample, as shown in the inset of **Figure 1a**. During the course of the fabrication, the MnBi$_2$Te$_4$ flake was exposed to the atmosphere for a short time only (see Fig. S1 for the details of the fabrications). The MnBi$_2$Te$_4$ device discussed in the main text is labeled as Sample 3 (S3). The height of S3 as measured by atomic force microscopy is ~7 nm (the inset of **Fig. 1c**), indicating it is a 5 septuple layers (SLs) device.

The temperature ($T$) -dependent resistance shows metallic behavior (Fig. 1a). The back-gate-voltage ($V_g$) -tuned resistance mapping for a range of temperatures from 2 to 40 K is shown in Fig. 1b. Bipolar behavior is shown for $V_g$ varying from -10 V to +60 V, and the charge neutral point (CNP) at low temperature is around +40 V. This implies that the band gap of a MnBi$_2$Te$_4$ device is achieved in this range of gate voltages, which provides us with the opportunity to observe the Chern insulator state. The resistance at the CNP ($R_{CNP}$) is extracted and shown in Fig. 1c. A a clear 'kink' around 22 K indicates this Néel temperature ($T_N$) of this sample. It is a little lower than that of the bulk crystal due to the increased thermal fluctuations seen in thinner flakes [16, 22]. The CNP would be shifted to a higher gate voltage with increasing temperature, especially where the Néel temperature is exceeded. Near the Néel temperature, $R_{CNP}$ is extremely large, at more than 100 kΩ. When $T$ > 22 K, $R_{CNP}$ exhibits insulating behavior, which can be described in terms of thermal activation.

The magnetoresistance for different gate voltages is shown in Fig. 1d. When $V_g$ = +55 V, the slope of the Hall resistance ($R_{xy}$) near zero field is negative, while it is positive when $V_g$ = +33 V. This implies that the carrier type changes from *n*- to *p*-type with decreasing gate voltage from +60 V to -10 V, consistent with the $V_g$-tuned resistance in Fig. 1b. The high-magnetic-field anomalous Hall effect (AHE) temains negative, independent of the carrier type. Specifically, the quantized anomalous Hall effect (also called the Chern insulator state) is observed when the gate voltage is around the CNP, accompanied by a near-zero longitudinal resistance ($R_{xx}$) when the magnetic field ($B$) is sufficiently strong to polarize the magnetic moment from an antiferromagnetic (AFM) to ferromagnetic (FM) state.

The magnetic field dependent transport is measured in order to characterize the Chern insulator state. The well-defined Chern insulator state is shown in **Figure 2c** when $B$ = 12 T and $T$ = 1.7 K, with $R_{xx} \sim 0.0002$ $h/e^2$ and $R_{xy} \sim -1.005$ $h/e^2$ (see Fig. S4). This indicates the precise quantization of the Chern insulator state. From the $V_g$ dependent resistance at 0 T and 12 T, we find that the CNP is in the quantization regime. This means that the quantization plateau retains its sign in both the $n$- and $p$-type regions, which rules out easily the possibility of QHE with LL. $R_{xx}$ and $R_{xy}$ mapping with the magnetic field and gate voltage (Figs. 2a, b) exhibits the quantization regime when $|B| \geq 6$ T. This magnetic field is the transition to the fully polarized FM state.

The temperature-dependent Chern insulator state is also shown. The quantization regime in the resistance mapping at $B$ = 12 T (Figs. 2d, e) shrinks clearly with increasing temperature. Below 22 K, the quantization value of $|R_{xy}|$ is larger than 0.97 $h/e^2$, and the longitudinal resistance minimum ($R_{xx\_min}$) is less than 0.04 $h/e^2$, as shown in Fig. 3a. This is adopted as an indication of the quantization criterion [3, 4, 16], and this temperature is the same as the Néel temperature of the sample. The Fermi-level-derived renormalized group (RG) flow [23] in the ($\sigma_{xy}$, $\sigma_{xx}$) space at 12 T and 2 K is shown in the inset of Fig. 2f. We find that it satisfies the semicircular scaling law (indicated by the dashed red line), which has been studied extensively [24]. The temperature-dependent converging point extracted from the RG flow is shown in Fig. 2f, which flows to (-1 $e^2/h$, 0 $e^2/h$) with decreasing temperature. This implies that the temperature- and Fermi-level-dependence of the RG flow provide a good description of the Chern insulator state.

**Validation of the dissipationless transport of the CES.**

The systematic temperature dependence of the dissipationless CES is studied. In **Figure 3a**, the temperature-dependent longitudinal resistance minimum and quantized anomalous Hall resistance in the quantization regime is shown. We note that the temperature can be divided into three regimes. When $T \leq 6$ K (regime I), $R_{xx\_min}$ is around 14 Ω, and it remains almost unchanged as indicated by the pink box in Fig. 3b. Such a small value of $R_{xx\_min}$ is a hallmark of the well-preserved dissipationless CES of a Chern insulator. When $T > 6$ K, $R_{xx\_min}$ increases sharply, which implies the breakdown of the dissipationless CES. When 6 K $< T <$ 22 K (regime II), $R_{xx\_min}$ changes from 36 to 1006 Ω. The linear relationship between $\ln T \cdot R_{xx\_min}$ and $T^{-1/2}$ in this temperature regime (the dashed red line in Fig. 3c) indicates that the transport behavior can be described by variable range hopping (VRH) [25-27]. The temperature-induced breakdown may thus be attributed to VRH. When $T > 22$ K (regime III), an Arrhenius plot with $R_{xx\_min} \propto \exp(-T_0/T)$ reveals the thermal activation behavior (Fig. 3d), where $T_0 = \Delta E/2k_B$, $\Delta E$ is the thermal activation gap, and $k_B$ is the Boltzmann constant. The gap extracted from the fit is $T_0 \sim 92$ K. Additionally, the transition temperature from VRH to thermal activation is also about the same as the Néel temperature of 22 K.

At the same time, the $R_{sh}$-$B$ curves in regime I reveal a critical magnetic field ($B_C$).

The four curves intersect at a point near critical magnetic field of -4.36 T (the inset of Fig. 3e). In the vicinity of $B_C$, we can plot the scaling behavior [17, 28-31] of $\ln R_{sh} \sim \ln(|B - B_C|/T^\kappa)$, as shown in Fig. 3e. Here, these curves would merge into one single curve, with the critical resistance being about 1.2 $h/e^2$ and the exponent parameter $\kappa \sim 0.35$. The two parameters are similar to those seen in other systems, such as the transition from QAH state to Anderson insulator [28], the transition from axion insulator to Chern insulator [17], the quantum Hall plateau transition [29], and so on. This finding implies that the transition here also shares the same universality class [32].

The current ($I$) -dependent transport has also attracted significant interest [33], especially the current-induced breakdown of the quantization [34]. At 2 K, the values of longitudinal resistance minimum in the quantization regime for different currents are shown in Fig. 3f. We find that $R_{xx\_min}$ stays around 14 $\Omega$ when the current is less than 1.4 μA, and starts to increase more obviously when $I > 1.4$ μA. Thus, the critical current ($I_C$) for the breakdown of the quantization can be identified as 1.4 μA at 2 K. The critical currents at different temperatures are shown in Fig. 3g, and the value decreases with increasing temperature. Furthermore, $I_C$ exhibits an almost linear relationship with temperature. Therefore, the effect of current on quantization may be similar to that of temperature. The current-induced breakdown of the dissipationless CES can be account for by two mechanisms, i.e., electric-field-driven VRH [35] and bootstrap electron heating [36, 37]. The breakdown that originates from bootstrap electron heating in QHE is often characterized by a large critical current (tens of μA) [38, 39], and there is usually an abrupt increas in the resistance [39, 40]. However, in this case the critical current for the Chern insulator state is relatively small (near 1 μA) here, and an abrupt jump is not seen. Therefore, the current-induced breakdown cannot be explained by electron heating in this case. In our work, the current-induced breakdown resistance is a few tens of Ohms (Fig. 3f), and it is as small as the resistance minimum in the VRH region. The small breakdown resistance, together with the linear relationship between the critical current and the temperature, suggests that the electric-field-driven VRH is the origin of the breakdown. The area shaded yellow in Fig. 3g thus marks the well-preserved dissipationless CES, and the part to the right of the dashed red line is the VRH region.

**Nonlocal measurement demonstrating the dissipationless CES.**

Nonlocal measurement with different probes is used to verify the dissipationless CES. The two-probe resistance $R_{12/12}$ (= $V_{12}/I_{12}$), three-probe resistance $R_{16/12}$ (= $V_{16}/I_{12}$), and four-probe resistance $R_{65/12}$ (= $V_{65}/I_{12}$) are shown against $B$ in **Figure 4a**. The inset indicates the numbers of the measurement probes. The two-probe resistance $R_{12/12}$ always exhibits a plateau of +1 $h/e^2$ at both positive and negative magnetic field when it settles into the quantization regime (here $V_g$ = 45 V and $T$ = 1.7 K), and the four-probe resistance $R_{65/12}$ is always near 0 $h/e^2$. It is the direct feature of the CES of a Chern insulator. Differently, the three-probe resistance $R_{16/12}$ is +1 $h/e^2$ in the negative magnetic field, and 0 $h/e^2$ in the positive field, which can be explained easily by the Landauer-Büttiker formalism[41, 42]. In a positive magnetic field, clockwise channel chirality is expected, and the voltage potentials are $V_1 = V_6 = V_5 = V_4 = V_3$, thus $V_{16}$ should be equal to $V_{65}$, which is zero. Therefore, $R_{16/12} = R_{65/12} = 0$ $h/e^2$. In a negative magnetic field, however, the voltage potentials are $V_6 = V_5 = V_4 = V_3 = V_2$, thus $V_{16}$ should be equal to $V_{12}$, showing $R_{16/12} = R_{12/12} = +1$ $h/e^2$. The chirality of the edge state is thus shown convincingly.

The temperature-dependent four-probe nonlocal resistance $R_{23/16}$ is shown in Fig. 4b. Like the longitudinal resistance in Fig. 2e, $R_{23/16}$ exhibits similar behavior. Values in the quantization regime ($R_{23/16\_min}$) at 12 T are extracted in order to confirm the dissipationless transport, and there are also three temperature regimes. In Fig. 4c, when the temperature is no more than 6 K, $R_{23/16\_min}$ is less than 5 Ω, suggesting well-preserved dissipationless CES. This can be explained by VRH between 6 K and 22 K, and by thermal activation over 22 K (in Figs. 4d, e). The thermal activation gap of the nonlocal measurement from the Arrhenius fitting is $T_{nl}$ ~ 85 K, while the nonlocal thermal activation gap is consistent with the local thermal activation gap $T_0$ ~ 92 K. In Fig. 4f, the plateau width $\Delta V_g$, defined as the difference between the peak and the valley in $dR/dV_g$ (the yellow circles in Fig. 4b), is extracted. At 2 K, the width can be as large as 15 V. It decreases with increasing temperature, and the nonlocal and local measurements exhibit the similar width. Consequently, nonlocal measurements reveal the dissipationless CES of the Chern insulator well, which has been widely studied in QHE and quantum spin Hall effect [43]. Both local and nonlocal measurements are effective methods for detecting the characteristics of the edge state.

**Conclusions**

In summary, we have observed the precise quantization of the Chern insulator state in the MnBi$_2$Te$_4$ device, with the well-defined quantization being maintained up to the Néel temperature of 22 K. Nonlocal measurements confirm the dissipationless CES of the Chern insulator. The transport behavior of the dissipationless CES was studied systematically by both local and nonlocal measurement. An increase in temperature increasing leads to three different regimes, i.e., well-preserved dissipationless CES, VRH and thermal activation. The increasing of the current can induce the breakdown of the well-preserved dissipationless CES, which is attributed to the electric-field-driven VRH. Our work reveals the novel scaling physics, and forms a complete set of evidences on the Chern insulator state in MnBi$_2$Te$_4$.

**Methods**

**Crystal Growth.** The MnBi$_2$Te$_4$ crystal was grown by the flux method. MnTe and Bi$_2$Te$_3$ were mixed in an alumina crucible with a molar ratio of 1:5.85, and then sealed in a quartz ampule. The mixture was heated to 950 °C over a period of one day. After maintaining it at a temperature of 950 °C for 12 h, it was cooled slowly to 580 °C. The MnBi$_2$Te$_4$ crystals were than obtained by centrifugation.

**Device Fabrication.** Thin flakes of MnBi$_2$Te$_4$ were exfoliated by tape, and then stamped on the 300-nm SiO$_2$ coated Si substrate. A few layers of graphite were transferred on to the MnBi$_2$Te$_4$ flake to protect the sample from the air and the photoresist. The graphite was transferred away prior to the electron beam lithography, and the gold electrode was then deposited. Following the lift-off process, the MnBi$_2$Te$_4$ field-effect-transistor was obtained. Finally, the h-BN was transferred on to the top of the MnBi$_2$Te$_4$ for protection. The transfer process was carried out in a glove box with the O$_2$ and H$_2$O each less than 0.1 ppm. Over the whole course of the fabrication, the MnBi$_2$Te$_4$ was exposed to air for a very short time (also see Fig. S1).

**Transport Measurement**. Electrical transport measurements were performed attemperatures down to 1.7 K and magnetic field up to 12 T. The voltage difference between different probes was measured using standard lock-in amplifiers (Stanford Research 830) with low frequency (13 Hz), and the back gate was applied using Keithley 2400 source meter. Except for the current-dependent transport shown in Figs. 3f, g, the current applied between the source and the drain was 100 nA. The longitudinal and Hall resistances are defined as $R_{xx} = V_{23}/I_{14}$ and $R_{xy} = V_{26}/I_{14}$, while the probe numbers are illustrated in the inset of Fig. 4a. The nonlocal resistance is defined in the main text, and the sheet resistance is given by $R_{sh} = R_{xx}(W/L)$, where $L$ and $W$ are the channel length and width, respectively. The conductivity is given by $\sigma_{xx} = R_{sh}/(R_{sh}^2 + R_{xy}^2)$ and $\sigma_{xy} = R_{xy}/(R_{sh}^2 + R_{xy}^2)$. The RG flow is plotted with $\sigma_{xy}$ on the abscissa and $\sigma_{xx}$ on the ordinate in ($\sigma_{xy}$, $\sigma_{xx}$) space.

**Acknowledgments**

We gratefully acknowledge the financial support of the National Key R&D Program of China (No. 2017YFA0303203), the National Natural Science Foundation of China (Grant Nos. U1732273, U1732159, 12025404, 12074181, 11904165, 11904166, 61822403, 11874203, 11834006, 91622115, 11522432, and 11574217), the Natural Science Foundation of Jiangsu Province (Grant Nos. BK20200007, BK20190286, BK20200312, and BK20200310,), the Fundamental Research Funds for the Central Universities (Nos. 021314380147, 020414380149, 020414380150, 020414380151, and 020414380152), the Users with Excellence Project of Hefei Science Center CAS (No. 2019HSC-UE007), and the opening Project of the Wuhan National High Magnetic Field Center.

**Author contributions**

Z.Y. fabricated the devices. Z.Y. and S.Z. performed the transport measurements. B.C. and F.F. grew the bulk crystals. S.Z., Z.Y. and F.S. wrote the manuscript with comments from all authors. F.S. was responsible for overall project planning and direction.

**Data Availability**

The data in this work are available from the corresponding author upon reasonable request.

**Figure Captions**

**Figure 1. Observation of the Chern insulator state in MnBi$_2$Te$_4$.**

**a** Temperature dependent resistance. The inset shows the optical graph of the device, and the white scale bar is 20 μm. **b** Gate voltage-dependent resistance mapping at various temperatures, showing the bipolar behavior with gate voltage. **c** The resistance of the charge neutral point ($R_{CNP}$) extracted from **a** exhibits the 'kink' around 22 K, which is the Néel temperature of the sample. The inset is the atomic force microscopy graph, and shows that the height of the sample is ~ 7 nm. The white scale bar is 1 μm. **d** The magneto-transport for various gate voltages when $T$ = 1.7 K. Near the charge neutral point, the quantized anomalous Hall effect is observed.

**Figure 2. Magnetic and temperature dependent chiral edge state.**

**a** $R_{xy}$ mapping with gate voltage and magnetic field at 1.7 K. **b** $R_{xx}$ mapping with gate voltage and magnetic field at 1.7 K. **c** Gate-dependent $R_{xy}$ and $R_{xx}$ at 1.7 K and 12 T. **d** $R_{xy}$ mapping with gate voltage and temperature at 12 T. **e** $R_{xx}$ mapping with gate voltage and temperature at 12 T. **f** Temperature-dependent converging points extracted from RG flow at 12 T. The inset shows the RG flow at 1.7 K and 12 T.

**Figure 3. Scaling of the dissipationless chiral edge state.**

**a** Temperature-dependent longitudinal resistance minimum ($R_{xx\_min}$) and quantized Hall resistance. **b** Well-preserved dissipationless CES below 6 K, with $R_{xx\_min}$ around 14 Ω. **c** VRH fitting. **d** Thermal activation fitting. **e** Scaling analysis in the vicinity of the critical magnetic field with the relationship $\ln R_{sh} \sim \ln(|B - B_C|/T^\kappa)$. The inset shows the critical magnetic field ~ 4.36 T. **f** Current-induced breakdown of the dissipationless CES at 2 K, with the critical current being ~1.4 μA. **g** Temperature-dependent critical current.

**Figure 4. Nonlocal measurement of the dissipationless chiral edge state.**

**a** Two-probe resistance $R_{12/12}$, three-probe resistance $R_{16/12}$, and four-probe resistance $R_{65/12}$ at 45 V and 1.7 K, from top to bottom. The inset defines the measurement probes numbers. **b** Nonlocal four-probe resistance $R_{23/16}$ mapping with gate voltage and temperature at 12 T. **c** Well-preserved dissipationless CES below 6 K. **d** VRH fitting. **e** Thermal activation fitting. **f** Plateau width at different temperatures.

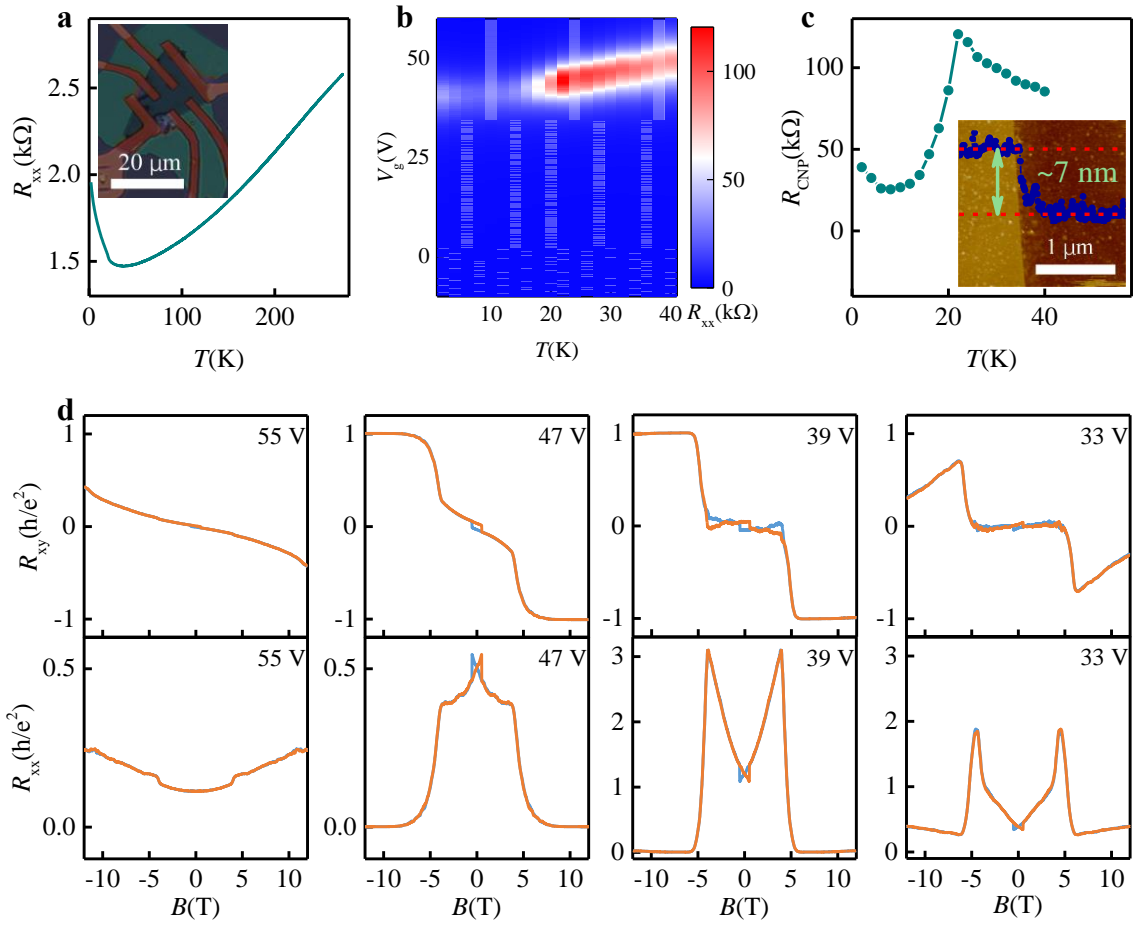

Figure 1

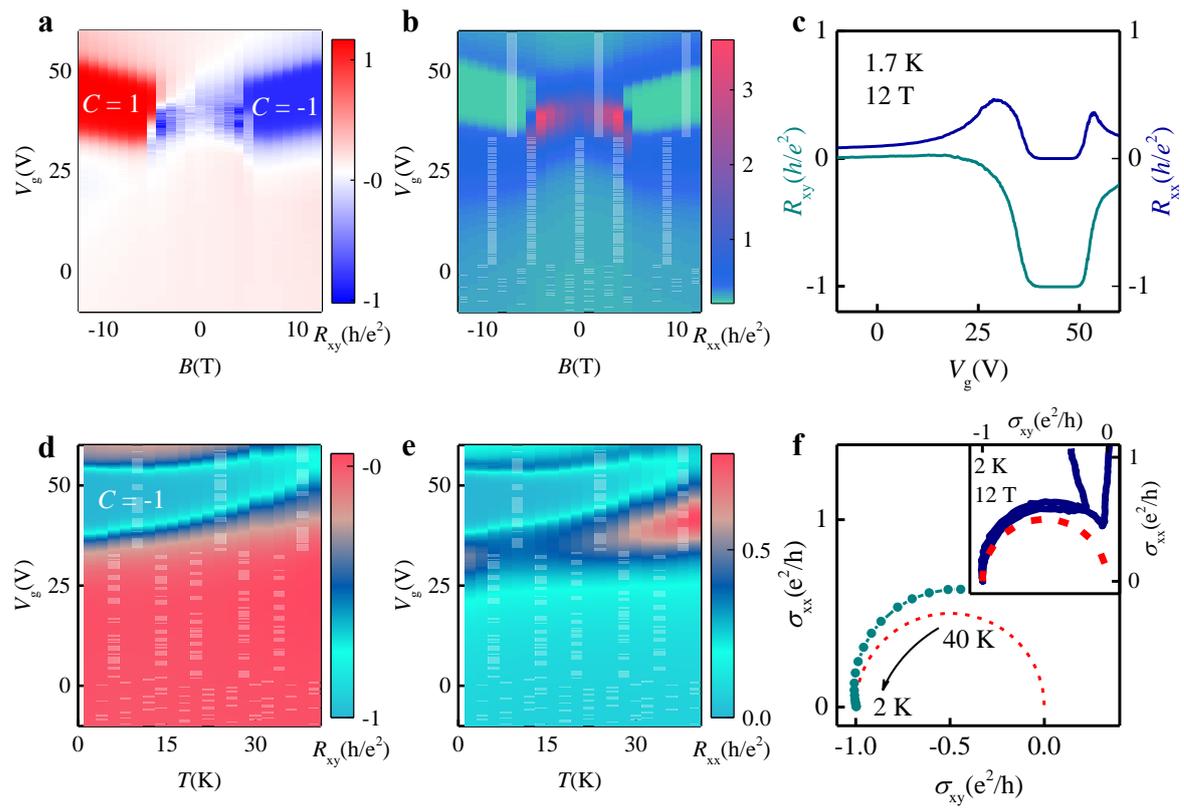

Figure 2

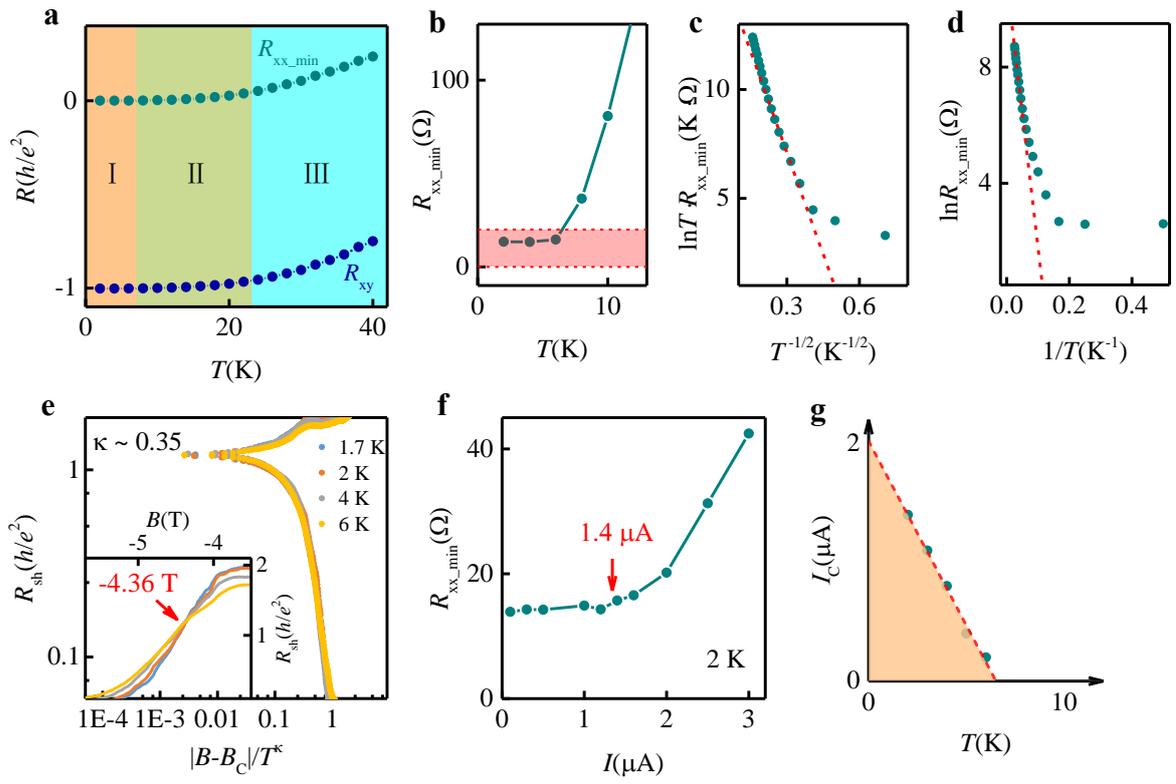

Figure 3

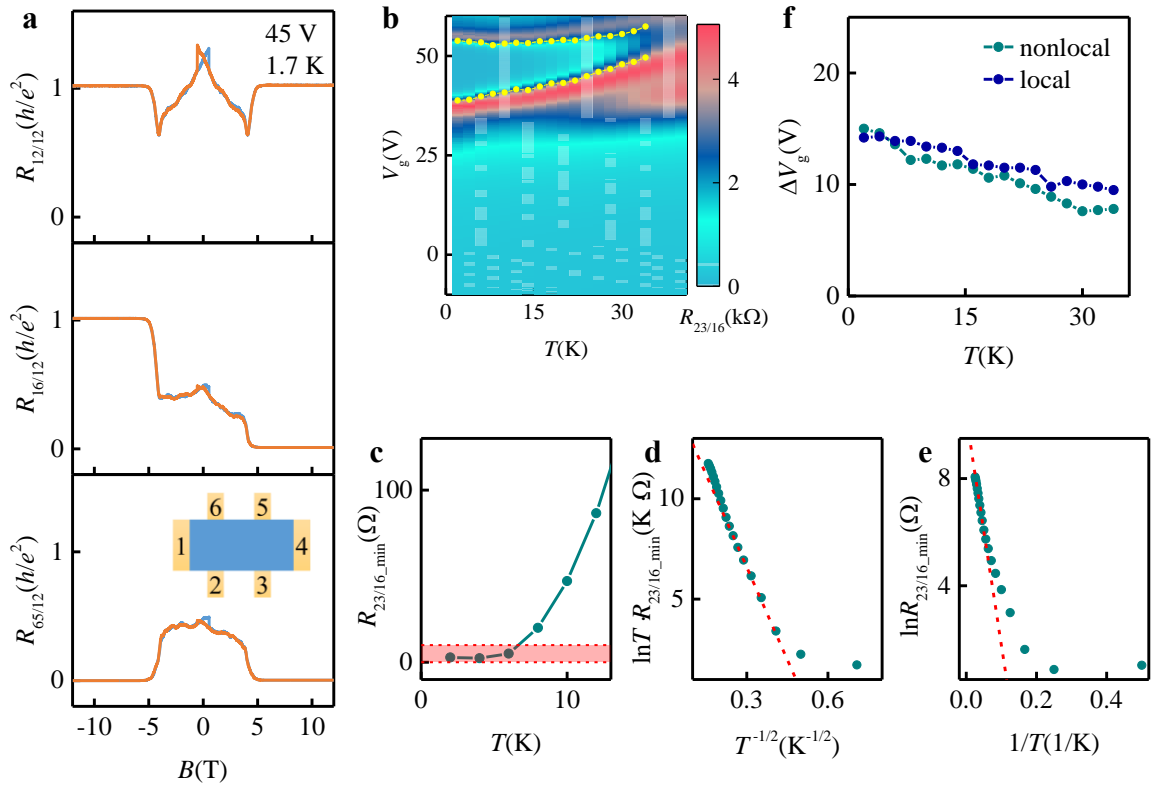

Figure 4